\documentclass[preprint, 12pt]{aastex}

\shorttitle{Interior Structure of Water Planets}
\shortauthors{Tian \& Stanley }

\begin{document}

\title{Interior Structure of Water Planets: Implications for their dynamo source regions}

\author{Bob Yunsheng Tian}
\affil{Department of Physics, University of Toronto, Canada}
\email{ytian@physics.utoronto.ca}

\author{Sabine Stanley}
\affil{Department of Physics, University of Toronto, Canada}

\begin{abstract}
Recent discoveries of water-rich, sub-Neptunian to Neptunian-massed exoplanets with short-period orbits present a new parameter space for the study of exoplanetary dynamos. We explore the geometry of the dynamo source region within this parameter space using 1-D interior structure models. We model planets with 4 chemically distinct layers that consist of (1) an iron core, (2) a silicate layer, (3) an H$_2$O layer, and (4) an H/He envelope. By varying the total planetary mass in the range of 1 - 19$M_{\oplus}$, the mass fraction of the H/He envelope between 0.1 - 5.1\%, and the equilibrium temperature between 100 - 1000 K, a survey of the parameter space for potential dynamo source region geometries is conducted. We find that due to the nature of the phase diagram of water at pressure and temperature conditions of planetary interiors, two different dynamo source region geometries are obtainable. Specifically, we find that smaller planets and planets with thicker H/He envelopes are likely to be in the regime of a thick-shelled dynamo. Massive planets and planets with thin H/He envelopes are likely to be in the regime of a thin-shelled dynamo. Also, small variations of these parameters can produce large interior structure differences. This implies the potential to constrain these parameters based on observations of a planet's magnetic field signature.
\end{abstract}


\section{Introduction}

With the growing number of extrasolar planets being discovered, a new population has emerged, bridging the gap between Earth-massed terrestrial planets and Neptune-massed giant planets. Compositional modeling of some of these planets based on their mass-radius relationship suggests an enrichment of heavy elements (C, N, O) \citep{Ada08, Rog10}, with metallicity more similar to Uranus and Neptune rather than to Jupiter or Saturn. Thus, like Uranus and Neptune, the interior of these bodies are thought to consist of a mixture of water (H$_2$O), ammonia (NH$_3$) and methane (CH$_4$) `ices' (the term `ices' is intended to specify the composition rather than a specific phase of the material). The class of water-rich, sub-Neptunian to Neptunian-massed `ice' giants with short-period orbits represents a new parameter space for many aspects of planetary science, including the study of planetary dynamos. 

In our solar system, the water-rich giant planets Uranus and Neptune exhibit non-axisymmetric, non-dipolar magnetic fields, in contrast to the axial-dipole dominated fields produced by other solar system dynamos. The geometry of the dynamo-generating layer of Uranus and Neptune likely explains this difference. Specifically, numerical simulations have demonstrated that a dynamo generated in a thin convecting shell of ionically conductive water phase surrounding a stably stratified layer can reproduce the unique magnetic field morphologies of the ice giants \citep{Sta04,Sta06}. This dynamo-region geometry was motivated by thermal evolution models of Uranus and Neptune used to explain their anomalously low intrinsic heat fluxes \citep{Hub95}. In contrast, the thick-shell geometries for the dynamo regions of Earth, Jupiter and Saturn appear to promote axial-dipolar dominated magnetic fields in these planets.

Recent research into phases of water at temperature and pressure conditions of giant planet interiors suggests that at higher pressures, exotic phase changes may occur from the ionic phase to either the plasma or the superionic phase depending on the temperature \citep{Cav99, Fre09, Net11}. Interestingly, interior models of Uranus and Neptune by Redmer et al. (2011) based on these new \textit{ab initio} equations of state and phase diagrams of water suggest a correspondence of the radius of the ionic-superionic phase transition and the radius of the stably stratified layers inferred from dynamo and thermal evolution models for these bodies.

The simple explanation for this correspondence would be that properties of the superionic phase result in an inhibition of convection (and hence dynamo action) in this region. This may result from a compositional gradient in the superionic layer, a 1st order increase in viscosity, or by a shallowing of the temperature gradient. In contrast to the 1st order ionic-superionic phase transition, the ionic-plasma phase transition is characterized by a gradual increase in conductivity between the phases \citep{Fre09}. It is therefore possible that this phase transition does not result in drastic changes in the stratification of the fluid. It should be noted that these predictions on the convective regime of plasma and superionic water phases have not been verified by either experiments or first-principle calculations, but they do seem plausible. Information on the thermal and compositional evolution of these bodies is required to determine if this assumption is valid.

In this paper, we use a 1-D interior model to determine the relative layer thicknesses of different water phases in Earth- to Neptune-massed water worlds. We study the effects of varying bulk mass, outer H/He envelope mass fraction, and equilibrium temperature (as a proxy for orbital radius and stellar luminosity). 1-D interior modeling of planets has been carried out for both purely rocky/icy bodies \citep{Val05, Val07, Sea07, Fu10} and bodies with a large gaseous envelope \citep{Ada08, Rog10, Net11} for a range of masses. Although different in some details, all of the studies assumed planets with differentiated layers of homogeneous and distinct composition, and then integrated the set of equations for hydrostatic equilibrium, mass conservation and the equations of state for each layer to obtain self-consistent interior models of the planets.

Motivated by the interior modeling studies of Redmer et al. (2011), in this study we assume that the dynamo does not operate in the superionic layer in extrasolar water worlds, but does operate in the combined ionic and plasma layers. Determining the relative thickness of these layers from our 1D models therefore determines the dynamo source region thickness. We use this parameter to distinguish water worlds that would produce Uranus or Neptune-like magnetic fields from those that would produce more Earth-like axially-dipolar dominated fields. We describe our interior structure model in Section 2. We present and discuss the results of the model runs in Section 3. In  Section 4 we discuss the implications for dynamo source regions, magnetic field morphologies and observables. Conclusions follow in Section 5.

\section{Model}
\subsection{Overview}

Our method for generating 1D interior structure models is similar to that of Rogers et al. (2011). We consider planets comprised of up to 4 distinct layers: (1) a metallic core layer, (2) a rocky layer, (3) an H$_2$O layer and (4) a hydrogen-helium envelope. It is possible to allow mixing of the hydrogen-helium envelope and the H$_2$O layer, with either a discrete or continuous increase in relative abundance of H$_2$O with depth. However, for the purpose of this study, the end-member case of chemically distinct layers is considered to reduce the number of free parameters.

In mass $m$ coordinates, we obtain profiles for radius $r(m)$,  pressure $P(m)$, density $\rho(m)$, and temperature $T(m)$ by integrating the equations of hydrostatic equilibrium and mass conservation:
\begin{equation}
\frac{dP}{dm} = -\frac{Gm}{4\pi r^4},
\end{equation}
\begin{equation}
\frac{dr}{dm} = \frac{1}{4\pi r^2\rho},
\end{equation}
completed by the thermal and caloric equations of state 
\begin{equation}
\rho = f(P,T),
\end{equation}
\begin{equation}
u = f(P,T),
\end{equation}
relevant to each layer, where $u$ is internal energy per unit mass and $G$ is the gravitational constant. The model inputs specify the mass of each layer $M_i$, the total mass of the planet $M_p$, the intrinsic luminosity $L_p$, and the equilibrium temperature of the planet. We employ the fourth-order Runge-Kutta method to integrate the equations for $r(m)$ and $P(m)$. We then calculate the temperature $T(m)$ accordingly. The equation of state is then used to determine the density $\rho(m)$. Using an initial guess for the planetary radius $r(m=M_p)$, we integrate equations (1) and (2) simultaneously inward towards the center of the planet. With each step inward, mass $dm$ is subtracted from the layer's mass $M_i$. Once $M_i  = 0$, we step to the next layer. Once we run out of mass, the radius $r(m=0)$ is used to calculate the next iteration's guess for $r(m=M_p)$. The model stops iterating when the condition $r(m=0) = 0$ is satisfied. 

We define the radius of the planet R as the radius at 1-bar pressure. The outer boundary conditions are as follows
\begin{equation}
r(M_p) = R,
\end{equation}
\begin{equation}
T^4(M_p) = T_{eq} ^4 + \frac{L_p}{4\pi R^2\sigma_B}.
\end{equation}
Here $\sigma_B$ is the Stefan-Boltzmann constant and $L_p$ is the luminosity due to a combined effect of secular cooling, envelope contraction, and radioactive heating. This parameter depends on the age and composition of the planet. For the purpose of our study, we keep $L_p$ a constant corresponding to the intrinsic luminosity of Neptune. Future studies on the effect of planet age can be done by varying this value, which has an effect on the radiative gradient and thus the depth at which the planet transitions into the adiabatic regime. However, we demonstrate in this study that this has very little effect on the deep interior structure of the body (see Section 3).

$T_{eq}$ is the equilibrium temperature of the planet, which depends on the luminosity of the parent star as well as the orbital distance by the following relation:
\begin{equation}
T_{eq}^4 = \frac{(1-A)L^*}{16\pi d^2\sigma_B},
\end{equation}
where $A$ is the Bond albedo of the planet, $L^*$ is the stellar luminosity, and $d$ is the average orbital distance. Figure 1 shows the variation of equilibrium temperature as a function of $L^*$ and $d$. The Bond albedo is kept constant at the typical value for solar-system giant planets of 0.33 \citep{Pea90}. The equilibrium temperature is not very sensitive to this choice of Bond albedo. For example, according to equation (7), if we allow the Bond albedo of the planet to be as low as that of a perfect absorber ($A=0$), the equilibrium temperature would be 11\% higher than the values on the contour in Figure 1. Conversely, if we let the Bond albedo to be as high as the value for Venus (0.90),  the equilibrium temperature would be 38\% lower.

\begin{center}
\begin{figure}
\center \includegraphics[trim = 50 150 50 250, clip, scale = 0.95]{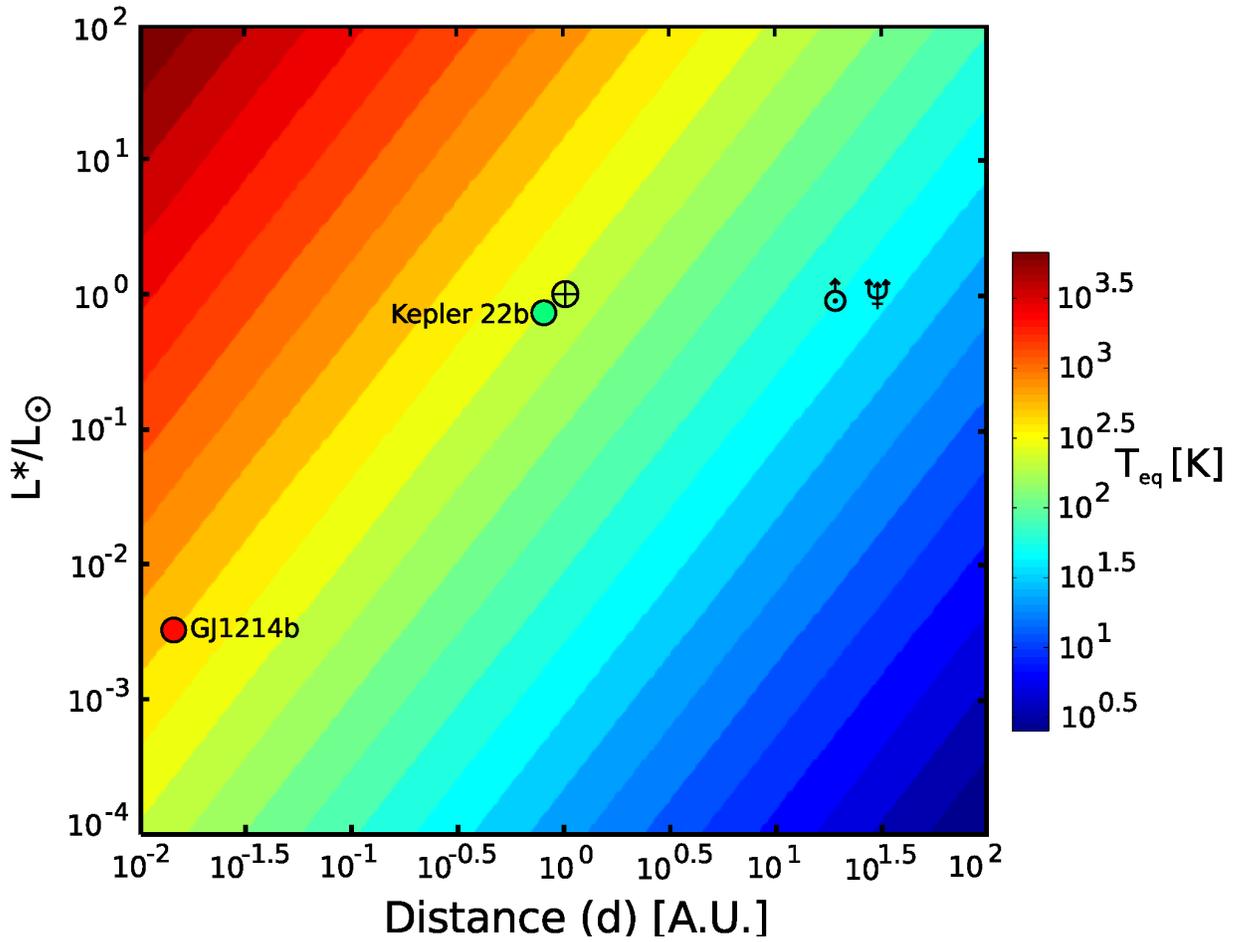}
\caption{The equilibrium temperature of a planet as a function of stellar luminosity (normalized by solar luminosity) and orbital distance. The values for Earth, Uranus, Neptune, and the purported water planets GJ1214b and Kepler 22b are also plotted.}
\label{fig:1}
\end{figure}
\end{center}

\subsection{H/He Layer}

In the outermost layer, we assume a gaseous envelope of H/He. Here, we use the Saumon-Chabrier-van Horn (SCvH) equation of state \citep{Sau95} to calculate the density and adiabatic temperature gradient ($\nabla_{ad}$). The actual temperature profile is calculated in three different regimes based on Rogers et al. (2010). In the outermost layers, the temperature profile is in a radiative equilibrium regime described by Hansen (2008) for irradiated atmospheres:
\begin{equation}
T^4 = \frac{3L_p}{16\pi R^2\sigma_B}\left[\tau+\frac{2}{3}\right]+\mu_0 T_{eq}^4\left[1+\frac{3}{2}\left(\frac{\mu_0}{\gamma}\right)^2-\frac{3}{2}\left(\frac{\mu_0}
{\gamma}\right)^3\ln\left(1+\frac{\gamma}{\mu_0}\right)-\frac{3}{4}\frac{\mu_0}{\gamma} e^{-\gamma \tau/\mu_0}\right].
\end{equation}
Here, $\tau$ is the infrared optical depth, $\gamma= 0.714\sqrt{T_{eq}/2000\mbox{K}}$ is the ratio between the optical and infrared optical depths, and $\mu_0$ is the angle cosine of incident radiation relative to surface normal, taken to be an average value of 1/2. The optical depth is calculated by integrating 
\begin{equation}
\frac{d\tau}{dm}=-\frac{\kappa}{4\pi r^2},
\end{equation}
where $\kappa$ is the opacity. The values for $\kappa(P,T)$ are taken from tables of Rosseland mean opacities of H/He gas with solar metallicity provided by Freedman et al. (2008). 
In the optically thick interiors ($\tau >> 1$), the planets transition to a radiative diffusion equation 
\begin{equation}
\nabla_{rad} \equiv \left(\frac{d\ln T}{d\ln P}\right)_{rad} = \frac{3\kappa L_p P}{64\pi\sigma_B GmT^4} .
\end{equation}
The radiative temperature gradient extends until the Schwarzschild criterion for convective stability:
\begin{equation}
\nabla_{ad} < \nabla_{rad}
\end{equation}
becomes violated, where
\begin{equation}
\nabla_{ad} \equiv \left(\frac{\partial\ln T}{\partial\ln P}\right)_s ,
\end{equation}
and subscript $s$ implies the derivative is taken at constant entropy. In this case, we transition into the final adiabatic regime, where the temperature profile is defined by the isentrope of the corresponding equations of state.

Isentropes are calculated using the provided caloric equation of state, which is obtained from a logarithmic interpolation of the quantum molecular dynamics (QMD) calculation of $u(P,T)$ by French et al. (2009). We write the isentropic thermodynamic relation as
\begin{equation}
ds = \frac{C_P}{T}dT - \frac{\alpha_V}{\rho}dP=0 ,
\end{equation}
where $\alpha_V$ is the volumetric thermal expansion coefficient:
\begin{equation}
\alpha_V\equiv \frac{1}{V}\left(\frac{\partial V}{\partial T}\right)_P ,
\end{equation}
and $C_p$ is the specific heat capacity at constant pressure:
\begin{equation}
C_P \equiv  \left(\frac{\partial h}{\partial T}\right)_P= \left(\frac{\partial u}{\partial T}\right)_P- \frac{P}{\rho^2}\left(\frac{\partial\rho}{\partial T}\right)_P .
\end{equation}
Here, $h$ is the enthalpy per unit mass. Substituting equations (14) and (15) into equation (13), we arrive at an expression for the isentropic temperature gradient in terms of internal energy per unit mass, density and pressure:
\begin{equation}
\left(\frac{\partial\ln T}{\partial\ln P}\right)_s = \left[1-\frac{\rho^2}{P}\left(\frac{\partial u}{\partial\rho}\right)_P\right]^{-1}.
\end{equation}

\subsection{H$_2$O Layer}

In the next layer, we assume the composition to be purely of water. In principle, an equation of state for a mixture of C, N, O, and H is needed to provide a representative interior structure model of these bodies (Nellis et al. 1997; Chau et al. 2010). However, due to the non-uniqueness in composition modeling of exoplanets, the relative abundances of C, N, O, and H in the mixture will be poorly constrained. This relative abundance will also vary for different planets. In order to properly account for the effect of variable composition of the C-N-O-H mixture layer, extensive QMD calculations of the equation of state at ultra-high pressure is required for each different C-N-O-H combination. Such calculation is computationally expensive, and introduces extra variables (i.e. ÔiceÕ layer composition) that are beyond the purpose of our study.

Such extensive QMD calculations have been conducted \citep{Fre09} and verified by experiments \citep{Knud12} for pure H$_2$O at pressures up to 700 GPa along the Hugoniot curve (densities up to 3.8 gcm$^{-3}$). When compared to `synthetic' Uranus calculations by Chau et al. (2010), which used a mixture with chemical abundance ratios of H:O:C:N=28:7:4:1, the phase diagram of pure H$_2$O and the C-N-O-H mixture showed similar phase boundaries. A comparison of the phase diagram of pure H$_2$O and the mixture is displayed in Figure 2. Additionally, H$_2$O is expected to be the most abundant material in ice-giants \citep{Hub95}. Thus, like many previous interior structure studies of Uranus and Neptune, we take H$_2$O as the end-member representative chemical for the `ice' layer in our models.

We use an interpolated version of the equation of state of H$_2$O obtained from QMD simulations by French et al. (2009) for applicable ranges of temperature and pressure. For pressures $<$ 0.709 kbar and temperatures $<$ 1000 K, a polytropic fit of a hybrid equation of state obtained by Seager et al. (2008) is used as an approximation. Like the H/He layer, the temperature profile is assumed to be adiabatic, and calculated from the isentropes of the equation of state.

\begin{center}
\begin{figure}
\center \includegraphics[trim = 75 100 50 200, clip, scale = 1]{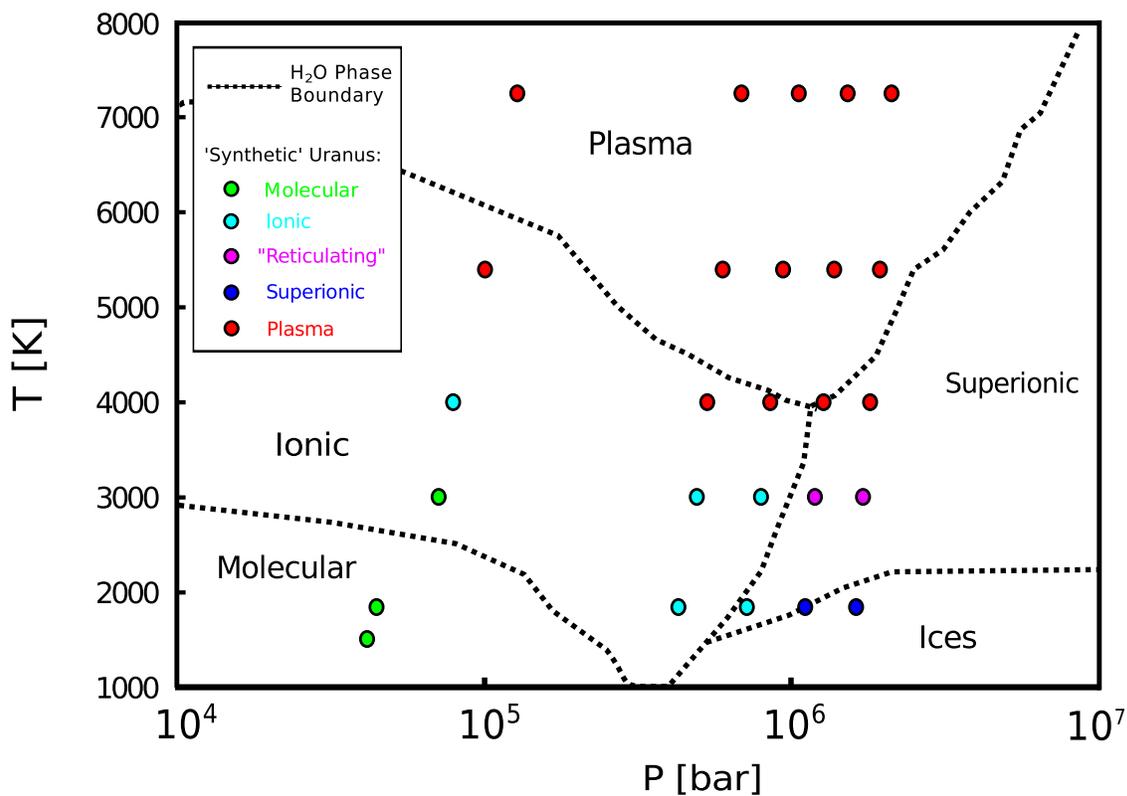}
\caption{The first-principle phase calculations (\textit{circles}) of a C-N-O-H mixture (with chemical abundance ratio of H:O:C:N=28:7:4:1) at several temperature and pressures from Chau et al. (2010), overlaying the phase diagram of pure H$_2$O (\textit{black dotted lines}) calculated by French et al. (2009). Colors of each point correspond to the phase of the C-N-O-H mixture, while the black text labels the phase of pure H$_2$O.}
\label{fig:2}
\end{figure}
\end{center}

\subsection{Silicate and Iron Layers}

The final two layers are composed of silicates and iron respectively. We use the polytropic fits of equations of state from Seager et al. (2008) to calculate the density of both layers. By using the polytropic equations of state, we neglect the fractional contribution of thermal pressure to the total pressure, and calculate density purely as a function of hydrostatic pressure. For justifications of this approximation and the quantification of thermal pressure, see Valencia et al. (2005) and Seager et al. (2008). Since the volume of the rocky core does not depend on the temperature in this approximation, the temperature profiles are not calculated in these layers.

\subsection{Parameters}

We explore the parameter regime of water-rich exoplanets in the mass range 1- $19M_{\oplus}$. Particularly, we look at the effects on the interior temperature vs. pressure profiles of the planet as we vary the mass, equilibrium temperature, and hydrogen-helium envelope mass fraction. The rocky/iron cores of the planets account for 10\% of the total planet mass in all models. We vary the bulk mass of the planet, the equilibrium temperature between 100 K - 1000 K, and the H/He envelope mass fraction between 0.1 - 5.1\%.

\section{Results}

\begin{center}
\begin{figure} 
\vspace{-3cm}
\center \includegraphics[trim = 20 20 20 20, clip, scale = 0.8]{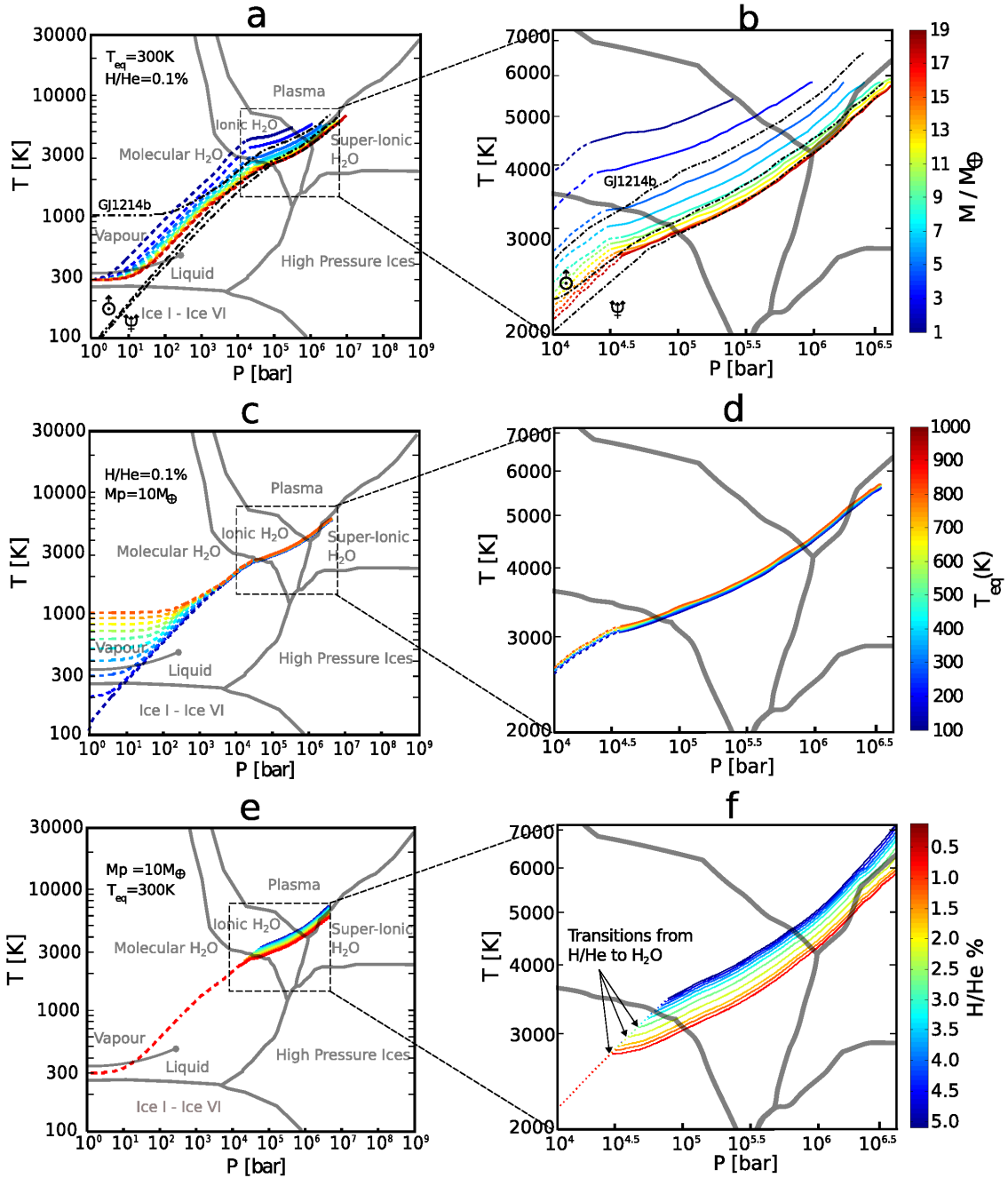}
\vspace{-2cm}
\caption{The temperature profiles in T-P space for planets with varying mass (a, b), $T_{eq}$ (c, d), and H/He mass fraction (e,f). Profiles for the Neptune, Uranus, and GJ1214b models are also displayed in (a) as black dash-dot lines. The phase diagrams are adapted from Nettelmann et al. (2011) for low pressures and French et al. (2009) and Redmer et al. (2011) for high pressures. Uranus and Neptune profiles are calculated using a 3-layer model allowing for H/He and H$_2$O in the top 2 layers. GJ1214b profile is modelled after Nettelmann et al. (2011) using 3 chemically distinct layers of H/He, H$_2$O and silicates. Dashed parts of the temperature profiles correspond to the hydrogen layer, and all of the profiles end at the outer boundary of the rocky cores.}
\label{fig:3}
\end{figure}
\end{center}

\subsection{Temperature Profiles}

The effects on interior temperature profiles as each of the parameters is varied are plotted in Figure 3, overlaying the phase diagram of water adapted from Redmer et al. (2011) and Nettelmann et al. (2011). We first  set the equilibrium temperature constant at 300 K, include a small H/He envelope accounting for 0.1\%  of the total mass, and vary the total planet mass (Figure 3a,b). We find that as mass increases between 1 - $19M_{\oplus}$, the temperature at a particular pressure decreases. Also, the higher-massed planets have an isentrope that can extend into the superionic and/or plasma water phases. 

We then vary the equilibrium temperature of the planet in Figure 3c,d, keeping the mass constant at $10M_{\oplus}$. Temperature profiles all converge to similar values in the deep interiors. It can be seen that the position at which the transition from radiative/isothermal to adiabatic regimes occurs is at a higher pressure when the temperature is higher. This is due to the fact that the radiative gradient is strongly temperature dependent (Eq. 10), while the adiabatic gradient is not. Thus, the efficiency of convective heat transport overcomes radiative transport at a higher pressure for hotter atmospheres. This result is also seen in previous studies using more sophisticated atmospheric models (Fig. 9 Fortney \& Nettelmann, 2009). It should also be noted that depending on $T_{eq}$ and $L_p$, the transition from radiative/isothermal to an adiabatic temperature profile may happen at P $<$ 1-bar, which is not captured by the model. However, for the parameter regime we studied, this only occurred for $T_{eq} = 100 K$.  

Since the majority of the planet's mass is in higher pressures, varying the equilibrium temperature should have very little effect on its interior temperature profile. However, depending on other parameters, some of the isentropes are so close to the plasma-ionic-superionic phase triple point, that even a small temperature difference in the deep interiors could lead to differences in the phase layer structures.

Finally, we keep mass and equilibrium temperature constant and vary the mass fraction of the H/He envelope between 0.1-5.1\%. Increasing the mass fraction of the H/He envelope heats the interior temperature profiles of the planet, but does not affect the temperatures in the outer hydrogen envelope. 

It should be noted that although each profile ends at the boundary between water and the silicate core, it starts at the outer boundary of the planet, rather than at the boundary between water and the H/He envelope. Therefore, a significant portion of the temperature profile is in the H/He layer and should not be referenced directly with the phase diagram of water given as the backdrop of Figure 3. We have marked the H/He portions of the temperature profiles with dashed colored lines in Figure 3, for which the background phase diagram would not apply.

The transition between the two chemically distinct layers can be best seen in Figure 3f. Here, since all of the profiles have the same $T_{eq}$ and mass, they follow the same P-T curve in the H/He envelope. The differences arise in the interior because of the depth at which the models transition into the H$_2$O layer, where the adiabatic gradient becomes shallower. Thus, planets with a higher \% of H/He would follow a steeper adiabatic gradient to higher pressures before the temperature profile reaches the H$_2$O layer and begins to shallow out, leading to a hotter interior. In Figure 3f, this transition happens at different points in the pressure range $10^4 - 10^5$ bar, and is visible as a `break' in the profile. This chemical transition can also be seen in Figure 3a, 3b. The pressure at which the change occurs increases slightly with increasing total mass, because as total mass increases, so does the total amount of H/He in the envelope.

These trends suggest that planets with lower equilibrium temperatures, higher masses or lower mass fractions of H/He envelopes will have lower temperature profiles in pressure-temperature space, which could lead to the presence of superionic phases in their interiors. Similar to previous studies, our models predict a thin ionic water layer for Uranus and Neptune possibly surrounding a superionic water layer \citep{Red11}, although the result depends on the poorly determined phase boundary between plasma and superionic phases.  In contrast, the higher equilibrium temperature of GJ1214b \citep{Net11} makes it less likely to reach the superionic phase.

\subsection{Thickness of the Dynamo Source Region}

As discussed in the introduction, we assume that the dynamo source region is the combination of ionic and plasma layers. We define the quantity $f$ as the ratio of the total thickness of dynamo generating layers to the outer radius of the dynamo source region. This gives us a proxy for whether the dynamo is operating in a thick-shelled or a thin-shelled regime. The values of $f$ are plotted in Figure 4, where each panel varies two of the three parameters.

The dynamo shell thickness $f$ increases gradually with increasing H/He mass fraction until a transition at 2.5\%, where it increases dramatically. Beyond 2.6\%, the shell thickness decreases with increasing H/He mass fraction gradually (Figure 4a). The marked change that occurs at around 2.5\% H/He results from the fact that temperature profiles for planets with higher H/He contents are raised enough that they no longer transition to the superionic phase (i.e. the temperature profiles remain in the plasma phase). This can be seen in Figure 3f where the temperature profiles corresponding to H/He fractions less than 2.5\% undergo a transition from plasma to superionic, whereas larger H/He fractions do not.

The value of $f$ decreases with increasing total mass (Figure 4b). It remains relatively constant with increasing equilibrium temperature (Figure 4c) for $T_{eq} > 200$ K. For example, the temperature profiles in Figure 3c,d all correspond to planets with a mass of 10 $M_{\oplus}$. This means they fall somewhere between the teal and green curves in Figure 4c. These curves do not exhibit much sensitivity of $f$ to $T_{eq}$, since the profiles in Figure 3c,d in the water regions are very similar and experience the phase transitions at very similar radii. 

Thus, the range of variation of $f$ in our study can be summarized by a contour plot of $f$ over our range of $M$ and H/He \% as displayed in Figure 5, where red regions correspond to a thick-shelled dynamo, and blue regions correspond to a thin-shelled dynamo.

\begin{figure}
\vspace{-0.5cm}
\center \includegraphics[scale=0.75]{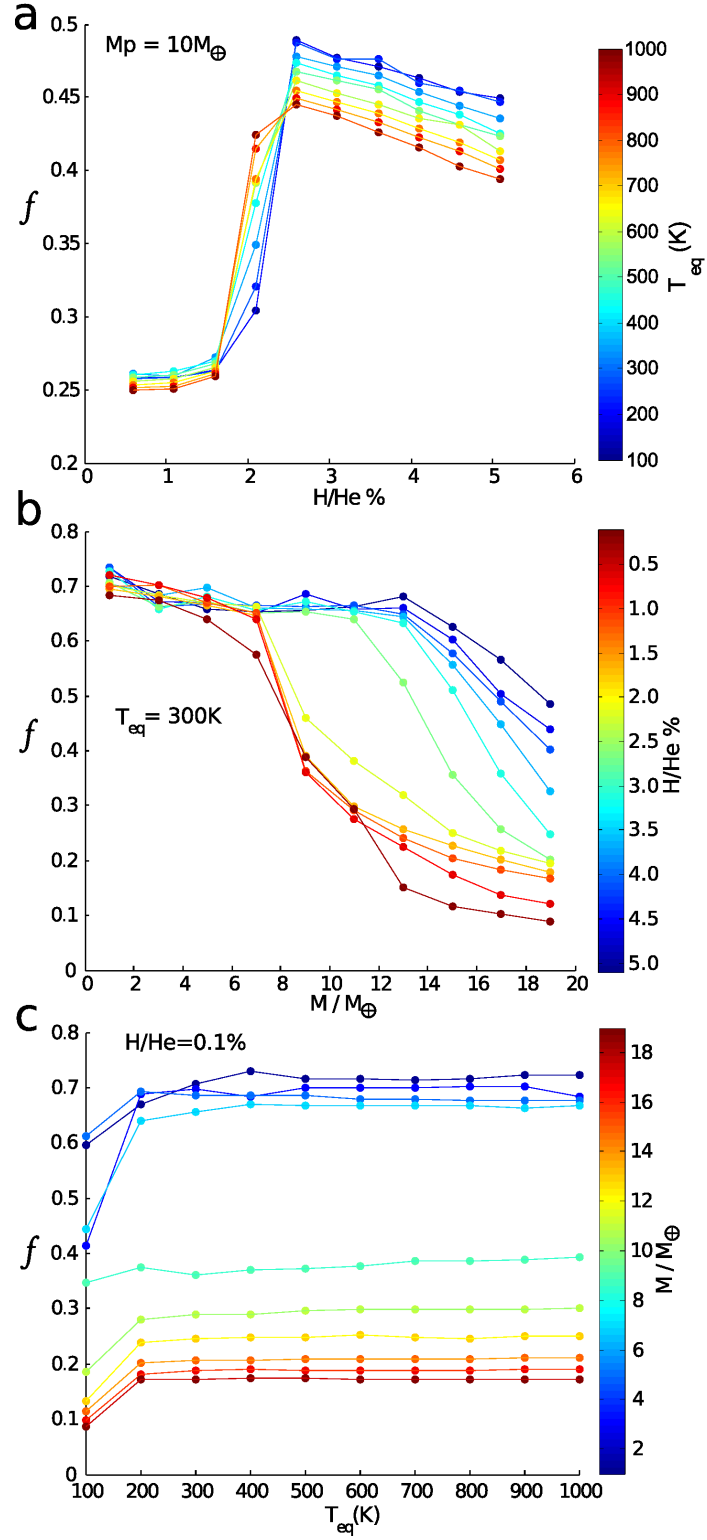} 
\vspace{-0.5cm}
\caption{Relative thickness of the dynamo generating layer to the total radius of the dynamo source region for models with constant mass (a), $T_{eq}$ (b), and H/He mass fraction (c).}
\label{fig:4}
\end{figure}

\begin{figure}
\center \includegraphics[trim = 100 200 100 200, clip, scale=1.1]{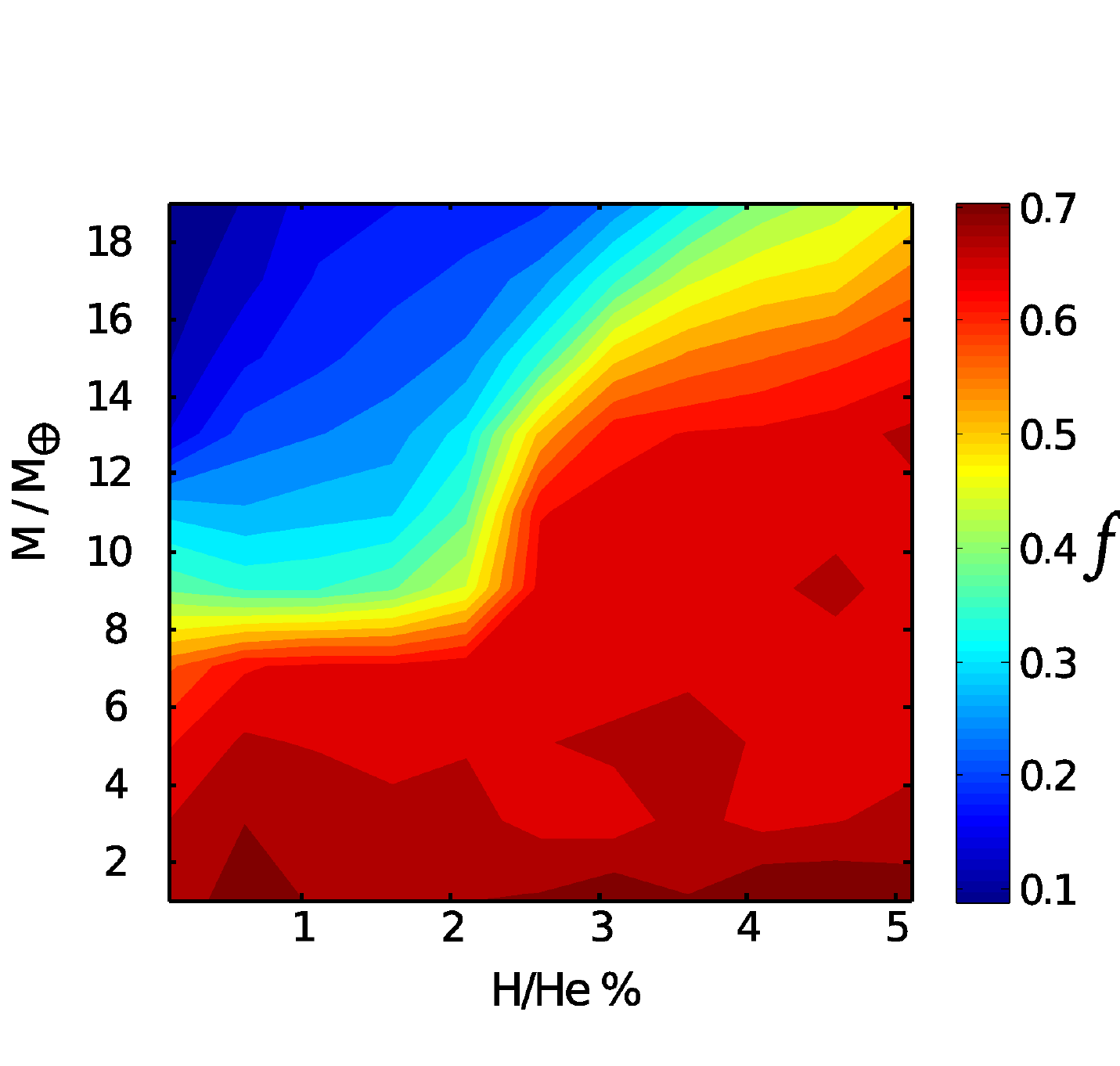}
\caption{Relative thickness of the dynamo generating layer to the total radius of the dynamo source region contoured over mass and H/He \%. $T_{eq}$ = 300 K for all runs.}
\label{fig:5}
\end{figure}

\subsection{Phase Layers: Varying $T_{eq}$ and H/He \%}

The thicknesses of the various water phase layers are displayed in Figure 6 for varying mass fraction of the H/He envelope, as well as for varying equilibrium temperature. The top plot shows that at $T_{eq}$= 300 K and $M_p = 10M_{\oplus}$, as we decrease the mass of the hydrogen envelope, the temperature profile is cooled enough that it crosses the plasma-superionic water phase interface shown in Figure 3f. This causes a regime shift in the interior structure to include a deep superionic water layer of considerable thickness. This regime change occurs when the H/He envelope is around 2.1 - 2.6\% of the total mass.

We study the impact of equilibrium temperature changes by varying this parameter between 100 - 1000 K. This was done at both 2.1 and 2.6\% H/He envelope mass fractions. As indicated previously, the equilibrium temperature only has a small impact on the interior structures. An increase in $T_{eq}$ thickens the H/He blanket relative to the total planetary radius. However, in this temperature range, for both H/He mass fractions, the relative thickness of the ionic and plasma water layers (and thus the geometry of the putative source region for the dynamo) remains only slightly affected by equilibrium temperature differences.

\begin{figure}
\vspace{-0.5cm}
\center \includegraphics[trim = 0 0 80 0, clip, scale=0.75]{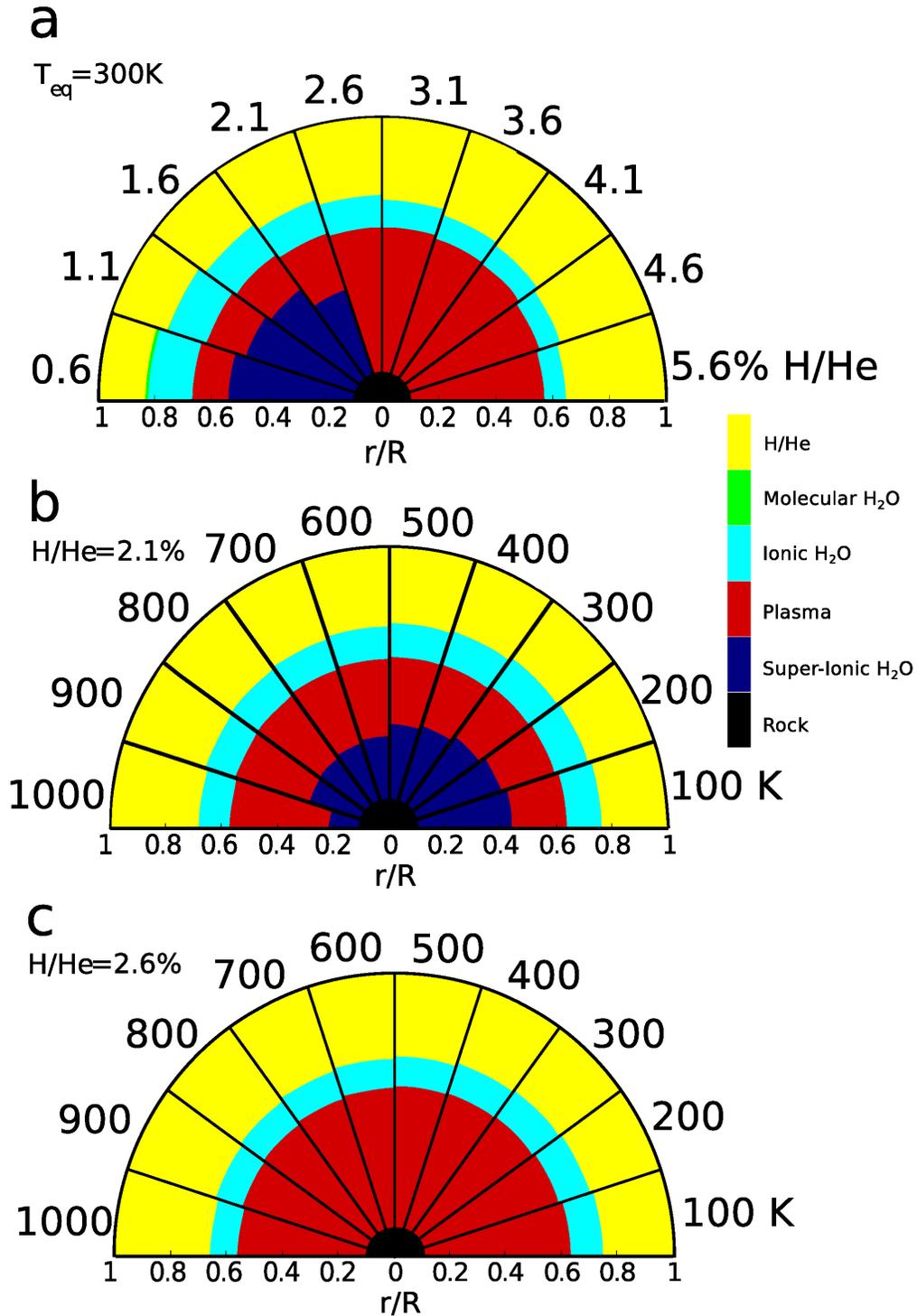}
\vspace{-1.5cm}
\caption{The interior structures of modelled planets with varying H/He mass fraction (a), varying $T_{eq}$ with H/He mass fraction held constant at 2.1\% (b) and 2.6\% (c). Each wedge is normalized  to its own radius.  This means that the actual planet radius varies between the models but is not represented in this figure.}
\label{fig:6}
\end{figure}

\subsection{Phase Layers: Varying $M_p$}
As expected, the total mass has a significant effect on the interior water phases present in the planets. Increasing the mass causes a thinning of the ionic water layer, and a thickening of the superionic water interior (Figure 7a). In this case, there is a regime shift that occurs when the equilibrium temperature is between 100 and 300 K. At low equilibrium temperatures, H$_2$O is separated into a very thin molecular H$_2$O layer, a shell of ionic H$_2$O, and a superionic interior. For the cases with higher equilibrium temperatures, there is a plasma phase that occurs between the superionic and ionic water layers. This layer of plasma thins as the total mass increases. Models that have $7M_{\oplus}$ and higher all have a significantly large core of superionic water with a thin shell of plasma and ionic water.

\begin{figure}
\vspace{-4cm}
\center \includegraphics[scale=0.8]{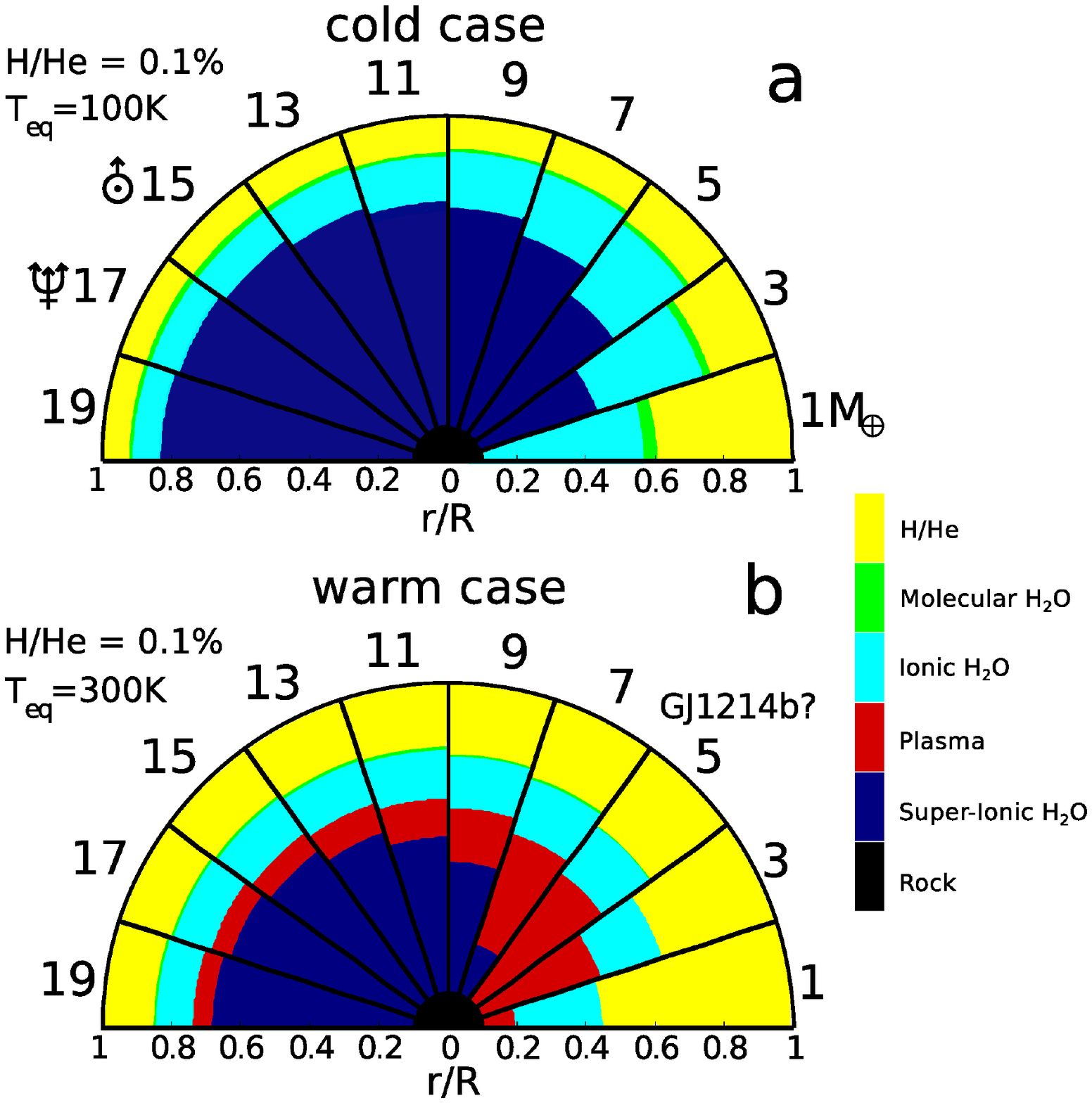}
\vspace{-4cm}
\caption{The interior structures of modelled planets with varying mass for H/He mass fraction held constant at 0.1\% and $T_{eq}$ held constant at 100 K (cold case) (a) and $T_{eq}$ held constant at 300 K (warm case) (b). For reference, we have indicated the wedges most similar to Uranus \& Neptune, and potentially, GJ1214b.}
\label{fig:7}
\end{figure}

\section{Discussion}
The strength and morphology of a planet's magnetic field are intimately tied to the planet's interior structure \citep{Sta04}. We have shown that by changing equilibrium temperature, mass, and H/He envelope mass fraction, a variety of interior structures can be achieved. There are other factors that can affect the planet's magnetic field, such as its thermal state, boundary conditions of the dynamo generating region, and the parameter regime (e.g. the Rossby number and the Ekman number). However, in this study we focus on the effects of the geometry.

In general, a thick-shell dynamo region geometry promotes more axial-dipolar fields, and a thin-shell geometry is more capable of generating non-axisymmetric, non-dipolar fields. However, the critical value of the geometry parameter $f$ at which the magnetic field transitions from an axial dipole to a non-axisymmetric, non-dipolar field will also depend on other factors such as the conductivity and viscosity profile of both the dynamo region and the non-convecting regions, as well as compositional gradients/boundaries within the convective region, which could produce layered convections and a double-dynamo regime \citep{Vil10}. All of these factors play a role in determining the surface magnetic field properties, and some can only be tested through full magnetohydrodynamics simulations.

It is possible that an iron core may also generate a dynamo, however, assuming it is relatively small (i.e. that we are dealing with planets that are predominantly ice-rich like Uranus and Neptune), it may be very difficult to observe any signature associated with a dynamo operating in this region.  This is due to the fact that power in the magnetic field decays with distance from the source.  The amplitude of decay is proportional to length-scale of the field, producing a ratio of surface field strength to core surface field strength of: 
\begin{equation}
\frac{B_{surface}}{B_{core}} = \left(\frac{r_{core}}{r_{surface}}\right)^{(l+2)} ,
\end{equation}
where $l$ is the spherical harmonic degree of the field length-scale being considered.  For example, the dipole component ($l=1$) at the surface will be weaker by a factor of $(r_{core}/r_{surface})^{3}$ compared to the value at the outer boundary of the core.  This means that unless the core field strength is very large, it is unlikely to be relevant for the observable external magnetic field.

\subsection{Dynamo Region Geometries in Cold Planets}
For planets with low equilibrium temperatures ($T_{eq} <$ 100 K) either due to being very far from their parent star or having a cooler parent star, the interior temperature profiles can be low enough to cross the ionic-superionic phase boundary (i.e. transition directly from the ionic to the superionic phase). For these planets, the majority of their interiors are dominated by ionic and/or superionic water phases regardless of the amount of hydrogen envelope present (at least in the range we consider). 

The geometry of the dynamo region in the cold regime depends on the mass of the planet. For low massed (1 - 2$M_{\oplus}$) objects, a thick-shelled dynamo region geometry occurs. As the mass increases, the dynamo eventually reaches a very thin shell geometry. Since thin shell geometry promotes non-dipolar, non-axisymmetric fields, the magnetic field morphology may change dramatically as a function of planetary mass.

\subsection{Dynamo Region Geometries in Warm Planets}
For planets with higher equilibrium temperatures ($T_{eq}>$  200 K), the interior temperature profile can be high enough to include thicker plasma layers either in place of or above the superionic water layers. 

For low massed to intermediate massed (1 - 5$M_{\oplus}$) objects, the dynamo source region has a thick-shell geometry. A dynamo in this regime would be more prone to generating an axial-dipolar magnetic field. For higher massed objects (6 - 19$M_{\oplus}$), the ionic and plasma layers thin. Below the plasma layer, a superionic water layer may exist depending on the thickness of the hydrogen envelope. The thinning of the dynamo source region may result in more non-dipolar, non-axisymmetric magnetic fields. 

For these warmer planets, the effects of an increased electrical conductivity with depth in the plasma layers may also result in specific magnetic field morphologies.  Dynamo models incorporating these features will be carried out in future work.

\subsection{Sensitivity to Parameters \& Phase Boundaries}
Since we have shown that the location of the dynamo region of the planet depends on its mass, equilibrium temperature, and H/He content, we can put constrains on these parameters based on observations of a planet's magnetic field (or \textit{vice versa}). Thus, it is important to quantify the sensitivity of temperature at high pressures to our parameters. To do so, we first define an equation that describes how the temperature at a particular pressure, which we take to be the approximate pressure of the plasma-ionic-superionic triple-point $T(P_1 = 100$ GPa$) = T_1$, changes when we change our three parameters in log-log space:
\begin{equation}
d\ln T_1 = \left(\frac{\partial\ln T_1}{\partial\ln T_{eq}}\right) d\ln T_{eq} + \left(\frac{\partial\ln T_1}{\partial\ln\chi}\right) d\ln\chi + \left(\frac{\partial\ln T_1}{\partial\ln M_p}\right) d\ln M_p,
\end{equation}
where $\chi$ is the H/He mass-fraction in units of \%. 

With this, we can write the \% change of $T_1$ as we vary our parameters in the following way:
\begin{equation}
\frac{dT_1}{T_1} = \frac{\left(\frac{\partial\ln T_1}{\partial\ln T_{eq}}\right)}{T_{eq}} dT_{eq} + \frac{\left(\frac{\partial\ln T_1}{\partial\ln\chi}\right)}{\chi} d\chi+ \frac{\left(\frac{\partial\ln T_1}{\partial\ln M_p}\right)}{M_p} dM_p,
\end{equation}
or
\begin{equation}
\frac{dT_1}{T_1} = \Theta_{T_{eq}} dT_{eq}+ \Theta_\chi d\chi+\Theta_M dM_p,
\end{equation}
Where, $\Theta_{T_{eq}}$, $\Theta_{\chi}$, $\Theta_{M}$ are our `sensitivity' coefficients that describe the \% change expected by $T_1$ when we vary our parameters by $dT_{eq}$, $d\chi$ and $dM_p$.

The sensitivity coefficient will change for various $M_p$, $\chi$ and $T_{eq}$. However, it is useful to quantify the range of values they can take on in the parameter space explored in our paper. For example, the values of $\Theta_{T_{eq}}$ for the parameter space explored (approximately 1000 models) ranges from 0.0002 - 0.002\%/K. This means that an increase of $T_{eq}$ by 100 K will raise the temperature at 100 GPa by only 0.2\% at most. This is seen in Figure 3, where we varied $T_{eq}$ by 900 K and did not raise the interior temperature by more than a few \%. Furthermore, the equilibrium temperature of an exoplanet is relatively well constrained by the stellar luminosity and orbital distance, and typically has error associated with the under constrained value for the Bond albedo. If we let the the Bond albedo of an exoplanet take on any value between 0 and 0.9 (see Section 2.1), the uncertainty in $T_{eq}$ is of the order 100 - 200 K, which would effect the calculation of interior temperature by only 0.2 - 0.4\%.

The range of values for $\Theta_{\chi}$ from our models are from 1 - 4\%/\%. The hydrogen envelope mass fraction is largely unconstrained by observations due to degeneracies in inverting for composition from the planet's mass-radius relationship \citep{Rog10}.  This can result in large differences of interior temperature. For example, if a planet has an H/He envelope accounting for 5$\pm$2.5\% of its total mass, its modelled deep interior temperature could have an error of up to 10\% or $\sim1000$ K. 

Finally, the range of values for $\Theta_{M}$ in our models are between -1 and -5 $\%/M_{\oplus}$. Depending on the detection method for the exoplanet, the uncertainties in mass can vary. If we allow for the uncertainty to be up to 2$M_{\oplus}$, it will lead to differences in interior temperature of up to 10\%.

In addition, the phase boundary between plasma and superionic phases used in this study has uncertainties due to the finite number of QMD simulations performed near the phase boundaries. Fig. 3 and Fig. 11 of French et al. (2009) show the distribution of simulation data points along the phase boundary. Since the phase of the layer is determined in our study by taking the phase of the closest data point in P-T space, our models are constrained by the P-T resolution of these data points. Given the density of data points along the phase boundary, the uncertainty of the phase boundary location ranges from 1 to roughly 5\%. In other words, at some places in P-T space, a plasma up to $\sim500$ K above the phase boundary we used could actually be in the superionic phase. 

This has implications for the critical H/He fraction and planet mass at which a regime shift from a thin-shelled to thick-shelled dynamo occurs in our models. Using the sensitivity coefficient calculated earlier, we determine that an uncertainty of 5\% in the position of the phase boundary translates to an uncertainty of 1 - 5\% in the critical regime shift H/He mass fraction. This means that the critical H/He \% could theoretically be somewhere between 0.1 and 7.5\% compared to the value of 2.5\% found in this study. An uncertainty of 5\% in phase boundary position also translates to an uncertainty of 1 - 5 M$_{\oplus}$ in the critical mass at which a planet changes dynamo regimes. Since interior temperature is so insensitive to the equilibrium temperature, extremely accurate determination of the phase boundary is required to say something useful about the effects of equilibrium temperature on dynamo regime in this parameter space. Thus, it is crucial to have more QMD calculations to more accurately determine the plasma-superionic phase boundary.

The phase boundary between molecular and ionic water also has uncertainties. However, since the transition between these two phases occurs at much lower pressure, shifting this boundary up and down will not greatly affect the spatial thickness of phase layers in the bulk of the interior. Thus, uncertainties associated with this phase boundary is not very relevant in calculations of $f$.

The effects of mixing H/He envelope with the H$_2$O layer is not explored in this paper. In principle, the entropy of mixing will lower the adiabatic gradient, as well as smooth out the `break' seen in the temperature profiles. The extent of this cooling effect cannot be fully quantified without an adequate equation of state for this mixture in various proportions. However, we do expect mixing to lower the temperature profiles, possibly into superionic phases. Thus, mixing could promote thin-shelled dynamo source regions.

The inclusion of ammonia (NH$_3$) and methane (CH$_4$) ices can also have an effect on the phase diagram of the model. According to Figure 2, the plasma-superionic phase boundary for a `synthetic Uranus' occurs about 1000 K lower than that for a pure H$_2$O composition. This suggests that an increased abundance of ammonia and methane would lower the phase boundary, which would have the same effect as raising the temperature profile, and hence results in a thicker plasma layer in the planet's interior. Thus, we expect the presence of methane and ammonia to promote thick-shelled dynamo source geometries and axial-dipolar external fields.

\subsection{Detectability of Magnetic Fields}
The interaction of a planet's magnetosphere with stellar wind from its parent star produces radio emissions. This is observed for all strongly magnetized solar system planets. The strength and frequency dependence of these radio emissions can be used to estimate the magnetic field strength. This was done successfully for Jupiter 20 years prior to the first \textit{in situ} magnetic field measurements \citep{Far99}.

The power of the radio emissions $P_{rad}$ is related to the magnetic dipole moment of the planet $M_{dipole}$ and its distance from the star $d$, such that:
\begin{equation}
P_{rad} \propto v^{7/3}\frac{{M_{dipole}}^{2/3}}{d^{4/3}} ,
\end{equation}
where $v$ is the velocity of the stellar wind particles \citep{Lec91}.

The power of the cyclotron radiation is also proportional to the `homogeneity' of the source $\Lambda$. This parameter depends on the average magnetic field gradient in the radial direction and the wavelength of the radio emission $\lambda$ \citep{Zar92}:
\begin{equation}
\Lambda = \left|\frac{\lambda}{R_p}\left<\frac{\partial\log B}{\partial r}\right> \right|^{-1}.
\end{equation}
Here, $<>$ denotes an average over the entire planet.

The sense of circular polarization of the radio signals can also indicate whether the auroral emissions are only coming from one pole of the magnetic field. Significant non-dipolar components and dipole tilt are also mechanisms that will modulate the radio emission frequency to the rotation period of the planet. Furthermore, the upper frequency limit of the radiation gives the maximum magnetic field strength near its surface \citep{Bas00}. Thus, the frequency spectrum and polarization of radio emissions can also give some information on the geometry of the magnetic field.

Thus, if it were possible to detect radio emissions from an exoplanet, one could infer some basic properties of field morphology and intensity which would allow some constraints on these planets' interiors based on the results of this paper.

\section{Conclusion}

We have shown that the nature of the H$_2$O phase diagram at planetary interior conditions allows for a variety of dynamo source region geometries to exist in the parameter space of Earth- to Neptunian-massed water-rich planets. Small variations in mass, equilibrium temperature, and composition can produce large differences in interior structure. These differences could manifest in the produced magnetic field morphologies from dynamos in these bodies. 

In this study, we found that small planets ($M < 5 \pm 2M_{\oplus}$) and planets with a high abundance of H/He ($\chi > 2.5\%$) are likely to be in the regime of a thick-shelled dynamo. Massive planets ($M > 10 \pm 2M_{\oplus}$) and planets with a low abundance of H/He ($0.1\% < \chi < 2.5\%$) are likely to be in the regime of a thin-shelled dynamo. Due to the high sensitivity of the interior temperatures on H/He mass fraction, the critical value for regime transition (found to be 2.5\% here) can be as high as 7.5\% and as low as 0.1\%. Furthermore, we expect the presence of methane and ammonia in the interiors to also promote a thick-shelled convective region. This difference in the thickness of the convective region can play a major role in deciding whether the planet has an Earth-like axisymmetric, dipolar magnetic field or a Neptune-like non-axisymmetric, non-dipolar field morphology. This is important for potential observations of exoplanetary magnetic fields  and their properties through the detection of radio emissions from these bodies. Such observation will also be a factor in the assessment of a planet's habitability.

The results presented above rely on accurate determinations of the plasma-superionic water phase boundary location. Thus, additional QMD simulations, to increase the resolution of the phase diagram in the region surrounding the triple-point and near the plasma-superionic phase boundary, may result in slightly different details regarding what specific parameter values produce thick- or thin-shelled dynamos (as seen in Section 4.3). Despite these uncertainties, the trend that hotter, smaller planets, and planets with thicker hydrogen/helium atmospheres will be more likely to avoid the superionic phase is robust. 

\section{Acknowledgements}

The authors thank two anonymous reviewers for useful comments that improved this manuscript. S. Stanley acknowledges funding for this project by the National Research Council of Canada (NSERC), and the Alfred P. Sloan Foundation.

\clearpage

\begin{thebibliography}{}
\bibitem[Adams et al.(2008)]{Ada08}  Adams, E. R., Seager, S., \& Elkins-Tanton, L.  2008, \apj, 673, 1160
\bibitem[Bastian et al.(2000)]{Bas00}  Bastian, T. S., Dulk, G.A., \& Leblanc, L.  2000, \apj, 545, 1058
\bibitem[Cavazzoni et al.(1999)]{Cav99} Cavazzoni, C., Chiarotti, G. L., Scandolo, S.,\& Tosatti, E., 1999, Science, 283, 44
\bibitem[Chau et al.(2010)]{Cha10} Chau, R., Hamel, S., \& Nellis, W. 2010, Nat. Comms., 2, 203
\bibitem[Farrell et al.(1999)]{Far99} Farrell, W. M., Desch, M. D., \& Zarka, P. 1999, \jgr, 104, 14025
\bibitem[Fortney \& Nettelmann (2009)]{For09} Fortney, J. J., \& Nettelmann, N. 2009, Space Sci. Rev. doi:10.1007/s11214-009-9582-x.
\bibitem[Freedman et al.(2008)]{Free08} Freedman, R. S., Marley, M. S., \& Lodders, K. 2008, ApJS, 174, 504
\bibitem[French et al.(2009)]{Fre09} French, M., Mattsson, T. R., Nettelmann, N., \&  Redmer, R. 2009, Phys. Rev. B, 79, 054107
\bibitem[French et al.(2010)]{Fre10} French, M., Mattsson, T. R., \&  Redmer, R. 2010, Phys. Rev. B, 82, 174108
\bibitem[Fu \& O'Connell (2010)]{Fu10} Fu, R. \& O'Connell, R. J. 2010, \apj, 708, 1326
\bibitem[Hansen(2008)]{Han08} Hansen, B. M. S. 2008, \apjs, 179, 484
\bibitem[Howard et al.(2010)]{How10} Howard, A. W., Marcy, G. W., Johnson, J. A., Fischer, D. A., Wright, J. T., Isaacson, H., Valenti, J. A., Anderson, J., Lin, D. N. C., \& Ida, S. 2010, Science, 330, 653
\bibitem[Hubbard et al.(1995)]{Hub95}  Hubbard, W. B., Podolak, M., \& Stevenson, D. J.  1995, Univ. of Arizona Press, Tucson, pp. 109-138
\bibitem[Knudson et al.(2012)]{Knud12} Knudson, M. D.,  Desjarlais, M. P., Lemke, R. W., Mattson, T. R., French, M., Nettelmann, N., Redmer, R. 2012, Phys. Rev. Lett., 108, 091102
\bibitem[Lecacheux(1991)]{Lec91} Lecacheux, A. 1991, Proceedings of Bioastronomy, the Search for Extraterrestrial Life - The Exploration Broadens, ed. J. Heidmann \& M.J. Klein (Berlin: Springer), 21
\bibitem[L\'eger et al.(2004)]{Leg04} L\'eger, A., Selsis, F., Sotin, C., Guillot, T., Despois, Mawet, D., Ollivier, M., Lab\`eque, A., Valette, C., Brachet, F., Chazelas, B., \& Lammer, H.  2004, Icarus, 169, 499
\bibitem[Mattsson \& Desjarlais(2006)]{Mat06} Mattsson, T. R. \&  Desjarlais, M. P. 2006, Phys. Rev. Lett., 97, 017801
\bibitem[Nellis et al.(1988)]{Nel88} Nellis, W. J., Hamilton, D. C., Holmes, N.C., Radousky, H. B., Ree, F.H., Mitchell, A.C., \& Nicol, M. 1988, Science, 240, 779
\bibitem[Nellis et al.(1997)]{Nel88} Nellis, W. J., Holmes, N.C., Mitchell, A.C., Hamilton, D. C., \& Nicol, M. 1997, \jcp, 107, 9096
\bibitem[Nettelmann et al.(2011)]{Net11} Nettelmann, N., Fortney, J. J., Kramm, U., \& Redmer, R. 2011, \apj, 733, 2
\bibitem[Pearl et al.(1990)]{Pea90} Pearl, J. C., Conrath, B. J., Hanel, R. A., Pirraglia, J. A., \& Coustenis, A. 1990, Icarus, 84, 12
\bibitem[Redmer et al.(2011)]{Red11} Redmer, R., Mattsson, T. R., Nettelmann, N.,\& French, M. 2011, Icarus, 211, 6
\bibitem[Rogers \& Seager(2010)]{Rog10} Rogers, L. A. \& Seager, S. 2010, \apj, 712, 974
\bibitem[Rogers et al.(2011)]{Rog11} Rogers, L. A., Bodenheimer,P., Lissauer, J., \& Seager, S.  2011, \apj, 738, 59
\bibitem[Saumon et al.(1995)]{Sau95} Saumon, D., Chabrier, G., \& van Horn, H. M. 1995, \apj, 99, 29
\bibitem[Seager et al.(2007)]{Sea07} Seager, S., Kuchner, M., Hier-Majumder, C., \& Militzer, B., 2007, \apjs, 669, 1279
\bibitem[Stanley \& Bloxham(2004)]{Sta04} Stanley, S. \& Bloxham, J. 2004, Nature, 428, 151
\bibitem[Stanley \& Bloxham(2006)]{Sta06} Stanley, S. \& Bloxham, J. 2006, Icarus, 184, 556
\bibitem[Valencia et al.(2005)]{Val05}  Valencia, D., Sasselov, D. D., \& O'Connell, R. J. 2005, Icarus, 181, 545
\bibitem[Valencia et al.(2007)]{Val07}  Valencia, D., Sasselov, D. D., \& O'Connell, R. J. 2007, \apj, 656, 545
\bibitem[Vilim et al.(2010)]{Vil10}  Vilim, R., Stanley, S., \& Hauck II, S. A.  2010, \jgr, 115, E11003
\bibitem[Zarka (1992)]{Zar92} Zarka, P. 1992, Adv. Space Res., 12, (8)99

\end{thebibliography}
\end{document}